\documentstyle[twocolumn,floats,aps]{revtex}

\begin{document}
\draft

\twocolumn[\hsize\textwidth%
\columnwidth\hsize\csname@twocolumnfalse\endcsname

\title{\bf Comment on "Dynamic correlations of the spinless Coulomb
Luttinger liquid [Phys. Rev. B {\bf 65}, 125109 (2002)]"}

\author{D. W. Wang$^{(1)}$, A. J. Millis$^{(2),(3)}$, and S. Das Sarma$^{(1)}$}

\address{
(1)Condensed Matter Theory Center, Department of Physics,
University of Maryland,
College Park, MD 20742\\
(2)Center for Materials Theory and Department of Physics and
Astronomy, Rutgers University, Piscataway, NJ 08554\\
(3)Department of Physics, Colombia University, New York, NY 10027
}

\date{\today}
\maketitle
\pagenumbering{arabic}

\begin{abstract}
We show that claims, presented in Phys. Rev. B {\bf 65}, 125109 (2002)
concerning the threshold behavior of dynamical correlations of a Coulomb 
Luttinger liquid, are based on an inconsistent mathematical analysis.
Physical arguments are also presented, indicating that the claimed behavior
is unlikely to be correct.
\end{abstract}

\vskip 1pc]
\narrowtext


In a recent paper \cite{gindikin02} Gindikin and Sablikov (GS) claimed that
remarkable logarithmic divergences occur in the charge density wave (CDW)
structure factor and the electron spectral function, $\rho(q,\omega)$, of one
dimensional spinless electronic systems with long-ranged Coulomb 
interactions. Their
results for the near-threshold behavior of the spectral function and
the CDW structure factor are respectively 
\begin{eqnarray}
\rho (q,\omega )\sim \frac{e^{-A(q)\beta |\ln \delta|^{1/2}}}{\omega\delta
|\ln \delta |^{1/2}}, \mbox{for}\ q\sim 0^+, 
\delta=\omega-\omega_q\to 0^+
\label{rho}
\end{eqnarray}
\begin{eqnarray}
S(q,\omega )\sim \frac{e^{-4\beta |\ln \epsilon|^{1/2}}}{\omega\epsilon
|\ln \epsilon |^{1/2}}, \mbox{for}\ q\sim 2k_F, 
\epsilon=\omega-\omega_{q-2k_F}\to 0^+
\label{S}
\end{eqnarray}
where wavevector $q$ and frequency $\omega$ are measured from Fermi
wavevector $k_F$ and Fermi energy $E_F$ respectively (we let $\hbar\equiv 1$
throughout this Comment); $\omega_{q}= 
qv(q)\sim v_F\beta^{-1} q\sqrt{-\ln(qd)}$
is the plasmon (boson) mode energy in the long wavelength limit
with $d$ being the characteristic length scale of the system.
In Eq. (\ref{rho}), $A(q)\equiv [v_F/v(q)-1]^2$,
and $\beta\equiv \lbrack \pi \hbar v_{F}/2e^{2}]^{1/2}$, where $v_{F}$ is Fermi
velocity and $v(p)$ is the renormalized plasmon (boson) velocity.
This remarkable result is argued to be a unique feature of the
Coulomb Luttinger liquid (CLL) and as GS note  is in disagreement with our
previous work \cite{wang00} which found that $\rho (q,\omega )$ rapidly
vanished as $\delta \rightarrow 0^{+}$.

In this Comment we show that the results of GS (Eqs. (\ref{rho}) and (\ref{S})
above) are incorrect. The error
arises from GS' incorrect assumption that their fundamental equation 
could be perturbatively expanded in a particular manner. We
show below that the expansion proposed by GS is not convergent; indeed
higher order terms neglected by them diverge more strongly near threshold
than the terms which they retain. We then provide a physical
argument further justifying the result originally presented in \cite{wang00}.

In order to clarify the notations, we 
first exhibit GS's main equation 
for the CDW structure factor  (Eq. (B1) 
of their paper) and then examine the convergence of the expansion
they have proposed. The electron spectral function may be
treated similarly (although we do not present explicit
results here) and is discussed  in the last portion of this Comment.

It is known
that the CDW dynamical structure factor is
proportional to the following function (note that we have shifted the
wavevector $q$ to be $q-2k_F$ for simplicity): 
\begin{eqnarray}
\tilde{F}(q,\omega)=\int^{+\infty}_{-\infty}dx\,e^{iqx}
\int^{+\infty}_{-\infty}dt\,e^{i\omega t}F(x,t),
\label{fourier_F}
\end{eqnarray}
where we use $"\,\,\tilde{}\,\,"$ to denote the Fourier component in
momentum-energy space, and
\begin{eqnarray}
F(x,t)&=&\exp\left(-v_F\int^{+\infty}_{-\infty}\frac{dp}{\omega_p}
(1-e^{-i\omega_p t-ipx})e^{-\alpha|p|}\right),
\label{f}
\end{eqnarray}
where $\alpha \rightarrow 0^{+}$ is a convergence parameter. Substituting Eq. 
(\ref{f}) into Eq. (\ref{fourier_F}), GS obtained their main equation by
integration by parts: 
\begin{equation}
\frac{\omega }{v_{F}}\tilde{F}(q,\omega )=\int_{-\infty }^{+\infty }dQ\tilde{
F}(Q,\omega -\omega _{q+Q})  
\label{main}
\end{equation}
where we have used the fact that $\tilde{F}(Q,\omega )$ is an even function
of $Q$. Eq. (\ref{main}) is the most important equation in Ref. \cite
{gindikin02} and is used to derive all of its important conclusions.

According to Ref. \cite{gindikin02}, the integrand of Eq. (\ref{main}) is
dominated by the leading term in a perturbative expansion of the {\it 
frequency argument} in powers of $Q$ about $Q=0$.
The authors of Ref. \cite{gindikin02} presented results only for 
the leading term; here we consider terms up to the second order of 
the expansion, and obtain
\begin{eqnarray}
&&\tilde{F}\left( Q,\omega -\omega _{q}-Q\omega _{q}^{\prime }-Q^{2}\omega
_{q}^{\prime \prime }/2\right)   \nonumber \\
&=&\tilde{F}(Q,\omega -\omega _{q})+\left[ -Q\omega _{q}^{\prime }-\frac{
Q^{2}}{2}\omega _{q}^{\prime \prime }\right] \frac{\partial \tilde{F}
(Q,\omega )}{\partial \omega }\Big\rfloor_{\omega =\epsilon }  \nonumber \\
&&+\frac{Q^{2}{\omega _{q}^{\prime }}^{2}}{2}\frac{\partial ^{2}\tilde{F}
(Q,\omega )}{\partial \omega ^{2}}\Big\rfloor_{\omega =\epsilon },
\label{exapnsion_F}
\end{eqnarray}
where $\omega _{q}^{\prime }$ and $\omega _{q}^{\prime \prime }$ are the
first and second derivative of $\omega _{q}$.
Since $\tilde{F}(Q,\omega )$ is an even 
function of $Q$, the linear term in $Q$ will not contribute to the
integration in Eq. (\ref{main}) so we obtain
\begin{eqnarray}
\frac{\epsilon +\omega _{q}}{2v_{F}}\tilde{F}(q,\epsilon +\omega
_{q})&=&\Phi_{1}(\varepsilon)+\Phi_{2}(\varepsilon)+\Phi_{3}(\varepsilon),
\label{exapnsion_F_int}
\end{eqnarray}
where
\begin{eqnarray} 
\Phi _{1}(\varepsilon ) &=&\int_{0}^{+\infty }
\tilde{F}(Q,\epsilon )\,dQ, 
\label{Phi_1}
\\
\Phi _{2}(\varepsilon ) &=&-\frac{\omega_q''}{2}
\int_{0}^{+\infty }dQ Q^{2}\frac{\partial \tilde{F}(Q,x)}{\partial x}
\Big\rfloor_{x=\epsilon }, 
\label{Phi_2}
\\
\Phi _{3}(\varepsilon ) &=&\frac{{\omega_q'}^2}{2}
\int_{0}^{+\infty }dQQ^2 \frac{\partial^{2}\tilde{F}(Q,x)}{\partial x^{2}}
\Big\rfloor_{x=\epsilon }.
\label{Phi_3}
\end{eqnarray}
In Eqs. (\ref{Phi_2}) and (\ref{Phi_3})
we have used a dummy variable, $x$, in the partial derivative to avoid
confusion. It is important to note that above expansion is not from the
total derivative of $\tilde{F}(Q,\omega -\omega _{q+Q})$ with respect to $Q$,
and it is valid only when $|Q^{\ast }\omega _{q}^{\prime }|$ is small
compared to both $\omega _{q}$ and $\omega -\omega _{q}$, where $Q^{\ast }$
is the maximum value of $Q$ in the integration and will be determined below.

Since it is known that $\tilde{F}(Q,\omega)$ is zero for $\omega<\omega_Q$,
the range of nonzero contribution in the integrand of Eqs. (\ref{Phi_1})-
(\ref{Phi_3}) is then determined by
\begin{eqnarray}
\epsilon=\omega-\omega_q \geq \omega_{Q}=\frac{v_F}{\beta} Q\sqrt{-\ln(Qd)},
\label{Q_max1}
\end{eqnarray}
from which we can obtain $Q^\ast$ in term of $\epsilon$ to the leading order
of small $\epsilon$:
\begin{eqnarray}
Q^\ast\sim\frac{\epsilon\beta}{v_F}\left[-\ln\left( \frac{\epsilon\beta d}{
v_F}\right)\right]^{-1/2} =\frac{\epsilon}{E_0 d}\left[-\ln\left( \frac{
\epsilon}{E_0}\right)\right]^{-1/2},  
\label{Q_max2}
\end{eqnarray}
where $E_0=v_F/\beta d$ is an energy scale. Therefore $|Q^\ast\omega_q^{
\prime}|$ can be always smaller than $\omega_q$ or $\epsilon$ as required for
the expansion, if $q$ is fixed to be a nonzero constant and $\epsilon$
is taken to be small enough.

In Ref. \cite{gindikin02}, GS neglected $\Phi _{2}(\epsilon)$ and 
$\Phi_3(\epsilon)$ in  Eq. (\ref{exapnsion_F_int})
and then used an elegant method to solve the resulting
equation $\frac{\epsilon +\omega _{q}}{2v_{F}}\tilde{F}(q,\epsilon +\omega
_{q})=\Phi _{1}(\varepsilon )$ directly. Their result, denoted as 
$\tilde{F}^{(1)}(q,\omega )$, is
\begin{eqnarray}
\tilde{F}^{(1)}(q,\omega)&\propto&\frac{2\pi v_F}{\omega}
\frac{e^{-4\beta\sqrt{-\ln((\omega-\omega_q)/E_0)}}}
{(\omega-\omega_q)\sqrt{-\ln((\omega-\omega_q)/E_0)}},
\label{F^1}
\end{eqnarray}
for $\omega -\omega _{q}=\epsilon $ much smaller than $E_{0}$. Note that $
\tilde{F}^{(1)}(q,\omega )$ shown above is divergent at the threshold, 
$\epsilon \rightarrow 0^{+}$.

To verify the consistency of GS's solution it is necessary to substitute
Eq. (\ref{F^1}) into the expressions for $\Phi _{2}(\epsilon)$ and 
$\Phi _{3}(\epsilon)$ and
verify that they are negligible relative to the leading order term. This
calculation  was not presented in Ref. \cite{gindikin02}. For simplicity we
analyze $\Phi _{3}(\epsilon)$ here by simple change of variables
($y\equiv x-\omega_Q$ and then $z\equiv \epsilon -\omega_{Q}$)
\begin{eqnarray}
\Phi_3(\epsilon)&=&\frac{{\omega_q'}^2}{2}
\int_0^{Q^\ast}dQQ^2\frac{\partial^2\tilde{F}^{(1)}(Q,y+\omega_Q)}
{\partial y^2}\Big\rfloor _{y=\epsilon-\omega_Q}
\nonumber\\
&\propto &\int_{0}^{Q^{\ast }}dQQ^{2}\frac{1}{
y+\omega _{Q}}\frac{e^{-4\beta \sqrt{-\ln (y/E_{0})}}}{y^{3}\sqrt{-\ln
(y/E_{0})}}\Big\rfloor_{y=\epsilon -\omega _{Q}}  \nonumber \\
&\propto &\int_{0}^{Q^{\ast }}dQQ^{2}\frac{1}{\epsilon }\frac{e^{-4\beta 
\sqrt{-\ln ((\epsilon -\omega _{Q})/E_{0})}}}{(\epsilon -\omega _{Q})^{3}
\sqrt{-\ln ((\epsilon -\omega _{Q})/E_{0})}}  \nonumber \\
&\propto &\frac{1}{\epsilon }\int_{0}^{\displaystyle\epsilon }\frac{
dz\,(\epsilon -z)^{2}}{\left[ -\ln ((\epsilon -z)/E_{0})\right] ^{3/2}}\frac{
e^{-4\beta \sqrt{-\ln (z/E_{0})}}}{z^{3}\sqrt{-\ln (z/E_{0})}},
\label{final}
\end{eqnarray}
where we have just kept the leading order term in small $y$ and $z$, because
the most important contribution of the $Q$ integration is from
$Q\sim Q^{\ast }$ for $\epsilon=\omega _{Q^{\ast }}$ \cite{gindikin02}. 
We see that the integration in 
Eq. (\ref{final}) is in fact strongly divergent, so the expansion proposed
by GS does not exist, and conclusions drawn from it 
(i.e. Eq. (\ref{S})) are not justified.

GS do not present a detailed calculation for the electron
spectral function, but assert that their result, Eq (\ref{rho}),
is obtained in a similar way to Eq (\ref{S}). We therefore
expect that their result is similarly invalid.
We now present a further argument that, independent of the details of any
particular analysis, the claim by GS that the spectral function of a Coulomb
Luttinger Liquid should diverge at threshold (i.e. Eq. (\ref{rho})), 
is not correct. We note that we
can study the ideal Coulomb Luttinger liquid by first introducing a
screening length in coordinate space and therefore a long wavelength
cut-off, $q_{0}$, in momentum space, for example 
\begin{eqnarray}
V(q)=2e^{2}|\ln ((|q|+q_{0})d)|\ \ \ \mbox{for }(|q|+q_{0})d\ll 1.
\label{screened_V}
\end{eqnarray}
The Coulomb Luttinger liquid is then approached by
taking the unscreened limit ($q_{0}\rightarrow 0^{+}$). In fact, we
note that essentially all physically relevant realizations of the Coulomb
Luttinger liquid are screened at some length scale, so that the limit $
q_{0}\rightarrow 0$ is of mainly mathematical interest.

According to the accepted  theory \cite{haldane,voit95},  the low energy
properties of a one dimensional electron gas with a short-ranged interaction
are determined by a bosonic Hamiltonian (Luttinger liquid model) with
exponent $\alpha $  given by the interaction strength at $q=0$ and a
threshold behavior of the spectral function given by \ $\rho (q,\omega
)\propto (\omega -\omega _{q})^{\alpha -1}.$ Applying this argument to the
screened Coulomb model yields  \cite{voit95,voit93}:
\begin{eqnarray}
\alpha  &=&\frac{1}{2}\left( \left[ 1+\frac{V(0)}{\pi \hbar v_{F}}\right]
^{-1/2}+\left[ 1+\frac{V(0)}{\pi \hbar v_{F}}\right] ^{1/2}-2\right)
\nonumber \\
&\sim &\frac{1}{2}\left( \beta ^{1/2}|\ln (q_{0}d)|^{-1/2}+\beta ^{-1/2}|\ln
(q_{0}d)|^{1/2}-2\right) ,
\end{eqnarray}
consistent with the proposal made in Ref. \cite{wang00}. As we approach the
ideal Coulomb Luttinger liquid by taking the limit $q_{0}\rightarrow 0^{+}$,
the effective exponent, $\alpha $, evidently becomes larger than one and
indeed ultimately diverges, and for $\alpha >1$ the standard Luttinger
liquid result predicts a spectral function vanishing at threshold. The
frequency dependence is concave, showing what Voit terms a "pseudo-gap"
structure \cite{wang00,voit93}. In addition, as explained in 
Ref. \cite{wang00}, the long-ranged interaction induces curvature in the
boson dispersion, further suppressing the threshold spectral weights,
$\rho(q,\omega\to\omega_q^+)$. This well known behavior is
inconsistent with the divergence at threshold proposed by GS. We also
emphasize that the results of effective exponents and "pseudo-gap" structure
\cite{wang00} are consistent with  results obtained via  numerical
calculation based on a direct Fourier transform from coordinate space
where electron Green's function can be expressed analytically by
bosonization method even for a long-ranged Coulomb interaction \cite{wang00},
and are also consistent to the results obtained in the renormalization group 
analysis \cite{RG}.

It is interesting to use similar effective exponent argument to study
the threshold behavior of the CDW structure factor. According to the
standard Luttinger liquid theory for short-ranged interaction \cite{voit95},
the CDW dynamical structure factor diverges at $\omega=\omega_{q-2k_F}$
with a power of $\alpha_{CDW}=-1+g(0)$ \cite{gindikin02,voit95}, where 
$g(p)=[1+V(p)/\pi\hbar v_F]^{-1/2}$. Using the screened short-ranged 
interaction shown in Eq. (\ref{screened_V}), $\alpha_{CDW}\sim
-1+\beta^{1/2}|\ln(q_0d)|^{-1/2}$ approaches to $-1$ logarithmically as 
$q_0\to 0$. If studying such threshold behavior by using the same method as in
Ref. \cite{wang00}, we expect to see a scale dependent effective exponent.
This confirms the existence of singularities in the CDW 
structure factor in the vicinity of $q=2k_F$ even for a long-ranged Coulomb
interaction. Therefore GS's criticism \cite{gindikin02} that our 
effective exponent \cite{wang00} will lead to a cusp threshold strucutre in
the CDW structure factor is misleading and incorrect. On the other hand, 
the correct divergent behavior should be 
not like Eq. (\ref{S}) as
proposed by GS due to their mistreatment toward the divergent expansion,
Eq. (\ref{exapnsion_F_int}), as discussed earlier. 

In summary, we find that the claims made in Ref. \cite{gindikin02}
concerning the threshold behavior of the CDW dynamical structure factor are
based on an inconsistent mathematical analysis. The spectral function
they obtained from the similar strategy is also shown to be 
in disagreement with
physical arguments and widely accepted results. 

This work is supported (DWW and SDS) by US-ONR, US-ARO, and DARPA, and by
NSF-DMR-00081075 (AJM). 


\begin{thebibliography}{7}
\bibitem{gindikin02} Y. Gindikin and V.A. Sablikov, Phys. Rev. B {\bf 65},
125109 (2002).

\bibitem{wang00} D.W. Wang, A.J. Millis, and S. Das Sarma, Phys. Rev. B {\bf
64}, 193307 (2001).

\bibitem{haldane} F. D. M. Haldane, J. Phys. C, {\bf 14}, 2585 (1981).

\bibitem{voit95} J. Voit,  Rep. Prog. Phys., {\bf 58}, 977 (1995).

\bibitem{voit93} J. Voit, J. Phys.: Condens. Matter {\bf 5}, 8305 (1993).

\bibitem{RG} S. Bellucci and J. Gonz$\acute{{\rm {a}}}$lez,  Eur. Phys. J. B
{\bf 18}, 3 (2000).

\end{thebibliography}
\end{document}